\documentstyle[multicol,aps,prl,psfig,array]{revtex} 

\renewcommand{\narrowtext}{\begin{multicols}{2} \global\columnwidth20.5pc}
\renewcommand{\widetext}{\end{multicols} \global\columnwidth42.5pc}
\def\bottom#1{\vskip #1\begin{picture}(290,80)(80,500)\thinlines \put(
330,500){\line( 1, 0){255}}\put(330,500){\line( 0, -1){
5}}\end{picture}}

\def\al{\alpha}
\def\be{\beta}
\def\ga{\gamma}
\def\de{\delta}
\def\ep{\epsilon}
\def\ve{\varepsilon}
\def\ze{\zeta}

\def\ka{\kappa}
\def\la{\lambda}

\def\rh{\rho}

\def\si{\sigma}

\def\ph{\phi}

\def\ch{\chi}
\def\ps{\psi}

\def\De{\Delta}

\def\mn{{\mu\nu}}

\def\cl{{\cal L}}

\def\fr#1#2{{{#1} \over {#2}}}
\def\frac#1#2{\textstyle{{{#1} \over {#2}}}}
\def\pt#1{\phantom{#1}}
\def\prt{\partial}

\def\half{{\textstyle{1\over 2}}}
\def\lsim{\mathrel{\rlap{\lower4pt\hbox{\hskip1pt$\sim$}}
    \raise1pt\hbox{$<$}}}
\def\gsim{\mathrel{\rlap{\lower4pt\hbox{\hskip1pt$\sim$}}
    \raise1pt\hbox{$>$}}}

\def\Re{\hbox{Re}\,}
\def\Im{\hbox{Im}\,}

\def\kaf{k_{AF}}
\def\kf{k_{F}}
\def\kfi{(k_{F})_{\ka\la\mu\nu}}
\def\kt{\tilde k}
\def\pht{\hat p}

\def\sd{\sin d} \def\cd{\cos d}
\def\sr{\sin r} \def\cor{\cos r}

\def\sx{\sin\xi} \def\cx{\cos\xi}
\def\sdp{\sin\de\ph} \def\cdp{\cos\de\ph}
\def\sxs{\sin^2\xi} \def\cxs{\cos^2\xi}

\def\etal {{\it et al.}}

\newcommand{\beq}{\begin{equation}}
\newcommand{\eeq}{\end{equation}}
\newcommand{\bea}{\begin{eqnarray}}
\newcommand{\eea}{\end{eqnarray}}
\newcommand{\rf}[1]{(\ref{#1})}

\begin{document}

\title{Cosmological Constraints on Lorentz Violation in Electrodynamics}     
\author{V.\ Alan Kosteleck\'y and Matthew Mewes}
\address{Physics Department, Indiana University, 
         Bloomington, IN 47405, U.S.A.}
\date{IUHET 438, July 2001, 
accepted for publication in Physical Review Letters} 

\maketitle

\begin{abstract}
Infrared, optical, and ultraviolet spectropolarimetry 
of cosmological sources is used to constrain
the pure electromagnetic sector 
of a general Lorentz-violating standard-model extension.
The coefficients for Lorentz violation
are bounded to less than $3 \times 10^{-32}$.
\end{abstract}


\narrowtext

Lorentz violation
is a promising candidate signal for Planck-scale physics
\cite{cpt98}.
For instance,
it could arise in string theory
\cite{kps}
and is a basic feature of noncommutative field theories
\cite{cga}.
In quantum field theory at attainable energies,
small violations can be incorporated into the standard model
to yield a general Lorentz-violating standard-model extension 
\cite{ck}.
Its lagrangian consists of all possible observer Lorentz scalars
formed from standard-model fields
while allowing for coupling coefficients with Lorentz indices.
All renormalizable and gauge invariant terms 
relevant at low energies are explicitly known.

The standard-model extension 
predicts definite experimental signals.
In the fermion sector of the theory,
various experiments have bounded coefficients for Lorentz violation.
However,
relatively little
is known experimentally about the implications
of the standard-model extension for the properties of light.
In particular,
no bounds have been placed on
the CPT-even coefficients for Lorentz violation in the photon sector.
In this work,
we study these terms and use
spectropolarimetry of cosmological sources
to obtain stringent bounds on Lorentz violation
comparable to the best current limits in the fermion sector.

Extracted from the standard-model extension,
the Lorentz-violating electrodynamics
can be written in terms of the usual field strength
$F_\mn \equiv \prt_\mu A_\nu -\prt_\nu A_\mu$,
which incorporates the electric field $\vec E$ 
and the magnetic field $\vec B$.
The relevant lagrangian terms are 
\cite{ck,fn1} 
\beq
\cl = -\frac 1 4 F_{\mu\nu}F^{\mu\nu}
- \frac 1 4 (\kf)_{\ka\la\mu\nu}F^{\ka\la}F^{\mu\nu}.
\label{lagrangian}
\eeq
The second term is CPT even and Lorentz violating.
Its coefficient
$\kfi$ is dimensionless.
It has the symmetries of the Riemann tensor
and zero double trace,
so it contains 19 independent real components.

The modified inhomogeneous Maxwell equations 
obtained from Eq.\ \rf{lagrangian} are
\beq
\prt_\al{F_\mu}^\al+(\kf)_{\mu\al\be\ga}\prt^\al F^{\be\ga} =0.
\label{max1}
\eeq
These are supplemented with the usual homogeneous Maxwell equations
$\prt_\mu \widetilde F^{\mn}=0$.
For a plane electromagnetic wave with wave 4-vector 
$p^\al = (p^0, \vec p)$,
we have 
$F_\mn (x) =F_\mn (p) e^{-ip_\al x^\al}$.
The homogeneous equations ensure the absence of single magnetic poles,
$\vec p \cdot \vec B =0$,
and provide the usual Faraday law,
$p^0\vec B = \vec p\times\vec E$.
This can be used to eliminate $\vec B$ 
in the modified inhomogeneous equations,
yielding the modified Amp\`ere law 
\cite{ck} 
\beq
M^{jk}E^k\equiv
(-\de^{jk} p^2 - p^j p^k -2(\kf)^{j \be \ga k} p_\be p_\ga) E^k
=0.
\label{ampere}
\eeq
The matrix $M^{jk}$ is real and symmetric,
which can be shown to imply lossless propagation.
The modified Coulomb law $p^jM^{jk}E^k=0$
follows from this equation,
in parallel with the usual case.

For nontrivial solutions to Amp\`ere's law,
we must require det$M^{jk}=0$. 
This condition provides the dispersion relation 
between $p^0$ and $\vec p$. 
Some calculation shows that,
to leading order in the coefficients $\kfi$,
the solutions to the dispersion relation take the form 
\beq
p^0_\pm=(1+\rh\pm\si)\left|\vec p\right|,
\label{kfdispersion}
\eeq
where
$\rh=-\half {\kt}_\al^{\pt{\al}\al}$
and $\si^2=\half(\kt_{\al\be})^2-\rh^2$,
with 
$\kt^{\al\be}\equiv (\kf)^{\al\mu\be\nu}{\pht}_\mu {\pht}_\nu$
and 
${\pht}^\mu \equiv {p^\mu}/{|\vec p|}$.
The solutions \rf{kfdispersion}
describe behavior similar 
to that of spatial dispersion in an optically inactive
but anisotropic medium.
In particular,
the vacuum is birefringent:
light typically propagates as two different independent modes. 
Substitution of the two solutions \rf{kfdispersion}
into the Amp\`ere law \rf{ampere}
determines the electric fields $\vec E_\pm$ 
and hence the polarization of the two modes.
For each mode,
the group velocity 
$\vec v_{g\pm}\equiv \vec\nabla_{\vec p}\ p^0_\pm$
typically differs from the phase velocity 
$\vec v_{p\pm}\equiv {p^0_\pm \vec p}/{\vec p^2}$.

At leading order in $\kfi$,
the fields $\vec E_\pm$ are orthogonal
and each is also perpendicular to 
$\vec v_{g\pm}$.
The orthogonality implies that $\vec E_\pm$ 
span the set of all possible polarizations,
and so the unit vectors 
$\hat\ve_\pm\equiv \vec E_\pm/|\vec E_\pm|$
form a basis for polarization at this order.
The electric field can then be decomposed as  
$\vec E(x) = 
(E_+ \hat\ve_+e^{-ip^0_+t}+
E_- \hat\ve_-e^{-ip^0_-t})
e^{i\vec p \cdot\vec x}$.
Since the phase velocities of the two modes differ,
their relative phase changes
as a wave propagates through free space.
The resulting change in the polarization state of the wave
is determined by the relative phase change
\beq
\De\ph =( p^0_+-p^0_-)t 
\approx 2\pi \De v_p L/\la \approx 4\pi\si L/\la,
\label{deph}
\eeq
where $\De v_p$ is the 
difference in phase velocities, 
$\la$ is the wavelength,
and $L$ is the distance the radiation has traveled. 
The distance dependence implies that,
for sources at very large distances,
tiny differences in the phase velocities may be observable.

In recent years,
high-quality spectropolarimetry of distant galaxies at 
infrared, optical, and ultraviolet frequencies has been performed
\cite{hough,brindle,cimatti465,dey,cimatti476,brothertonfirst,brotherton,%
brothertonqso,kishimoto,vernet}.
These observations 
correspond to $L/\la$ greater than $10^{31}$.
We can therefore anticipate 
that measurements of polarization parameters of order 1
should provide sensitivity of order $10^{-31}$ or better
to components of $\kfi$.
The inverse dependence of $\De \ph$ 
on wavelength is a special feature
exploited here to obtain a definite bound on $\kfi$.

In spectropolarimetry,
the quantitative measurement of polarization 
is usually described via Stokes parameters,
defined in a frame where the 3-axis 
coincides with the direction of propagation 
\cite{bw}.
Introducing unit vectors $\hat e_1$ and $\hat e_2$
along the 1- and 2-axes
and the corresponding electric field 
components as $E_1$ and $E_2$,
the Stokes parameters can be taken as 
\bea
(s^0,\vec s)&\equiv&
\bigl( |E_1|^2+|E_2|^2,~ |E_1|^2-|E_2|^2,~
\nonumber\\
&&\qquad\qquad 2\Re{{E_1}^*E_2},~ 2\Im{{E_1}^*E_2}\bigr)
\nonumber\\
&=&
s^0 ( 1,~ \cos2\ch\cos2\ps,~ \cos2\ch\sin2\ps,~ \sin2\ch),
\eea
where $\ch$ and $\ps$ are the usual polarization angles.
For convenience, we normalize throughout to $s^0=1$.
In the present context,
nonzero coefficients $\kfi$ leading to a nonzero $\si$ 
imply a finite phase shift $\De\ph$
in the radiation from a cosmological source,
which in turn affects the Stokes vector $\vec s(\ps,\ch)$
determining its polarization.
To set a bound on Lorentz violation,
we must therefore first establish quantitatively the relationships
between $\kfi$, $\si$, $\De\ph$, and $\vec s(\ps,\ch)$.

We begin our analysis by expressing $\si$ directly  
using the 19 independent components of $\kfi$.
Since $\si^2$ is a quadratic form in $\kfi$,
we can choose 19 independent components 
$k_A$, $A,B=1, \dots, 19$, of $\kfi$ 
and can write $\si^2=S_{AB}k_Ak_B$.
The $19\times 19$ matrix $S_{AB}$ 
is symmetric and depends on the direction of propagation.
Some calculation shows that 
there exists a direction-independent similarity transformation 
such that $S_{AB}$ takes the form of a $19\times 19$
matrix with only a $10\times 10$ nonzero block.
Therefore,
$S_{AB}$ has rank 10, and 
only ten linearly independent combinations 
$k^a$, $a = 1, \ldots, 10$,
of $\kfi$ appear in $\si$.
An acceptable choice for these ten combinations is 
\bea
k^a &=& \bigl( 
(\kf)^{0213},~
(\kf)^{0123},~
\nonumber\\
&&\qquad
(\kf)^{0202}-(\kf)^{1313},~
(\kf)^{0303}-(\kf)^{1212},~
\nonumber\\
&&\qquad
(\kf)^{0102}+(\kf)^{1323},~
(\kf)^{0103}-(\kf)^{1223},~
\nonumber\\
&&\qquad
(\kf)^{0203}+(\kf)^{1213},~
(\kf)^{0112}+(\kf)^{0323},~
\nonumber\\
&&\qquad
(\kf)^{0113}-(\kf)^{0223},~
(\kf)^{0212}-(\kf)^{0313}
\bigr).
\label{ka}
\eea
It now follows that $\si^2=\tilde S_{ab}k^ak^b$,
where $\tilde S_{ab}$ is symmetric and direction dependent.
The other nine linearly independent combinations of $\kfi$
play no role in birefringence,
and bounding them will be the 
subject of a separate investigation.
For definiteness,
the reference inertial frame
in which the $k^a$ are specified by Eq.\ \rf{ka} 
is chosen to be compatible with celestial equatorial coordinates,
with the 3-axis aligned along 
the celestial north pole at equinox 2000.0
at a declination of $90^\circ$.
The 1- and 2-axis are at a declination of $0^\circ$
and a right ascension of $0^\circ$ and $90^\circ$,
respectively.
The goal is to bound the ten quantities $k^a$ 
defined in this frame.

The form of $\tilde S_{ab}$ is cumbersome
and is omitted here.
A more convenient expression for $\si^2$ 
can be obtained by calculating within a
special inertial frame.
The idea is to use observer rotation invariance 
to choose a `primed' frame 
in which $\hat p^\mu$ has leading-order form
$\pht\, '^{\mu}=(1;0,0,1)$,
to solve for the Lorentz scalar $\si^2$ in this frame 
in terms of $\kfi '$, 
and then to use the rotation 
between the celestial equatorial frame and the primed frame
to express $\kfi '$ in terms of $\kfi$.
To match standard polarimetric conventions, 
we choose the primed-frame basis vector $\hat e_3'$ to point
from the source towards the Earth, 
while $\hat e_1'$ points south
\cite{fn2}.

Solving the Amp\`ere law \rf{ampere} in the primed frame gives 
$\rh=\half(\kt '^{\, 11}+\kt '^{\, 22})$
and
$\si^2=(\kt '^{\, 12})^2
+\fr 1 4 (\kt '^{\, 11}-\kt '^{\, 22})^2$.
This form for $\si^2$ suggests defining an angle $\xi$ such that 
$\kt '^{\, 12} = \si \sin\xi$ and 
$\half(\kt '^{\, 11}-\kt '^{\, 22})= \si \cos\xi$.
The angle $\xi$ determines
the minimal linear combinations of $\kfi$
relevant for polarimetry of a specific source.
Note that $\xi$ is frame dependent,
unlike $\rh$ and $\si$.

At leading order,
the polarization basis vectors in this frame obey 
$\hat\ve_\pm\propto(\sin\xi,\pm1-\cos\xi,0)$
and are linearly polarized.
The corresponding Stokes vectors are
$\vec s_\pm =\pm (\cos\xi,\sin\xi,0)$.
The propagation from the source to the Earth 
generates a relative phase change $\De\ph$
specified by Eq.\ \rf{deph}.
The corresponding effect on the Stokes vector $\vec s(\ps,\ch)$
for the radiation can be regarded as a rotation by $\De\ph$
about the basis vector $\vec s_+$.
This typically changes both $\ps$ and $\ch$.

The change in polarization depends
not only on the coefficients $\kfi$, 
the wavelength $\la$, and the distance to the source $L$,
but also on the initial polarization.
For cosmological sources,
there is usually no way to  
determine independently the polarization produced at the source.
We adopt instead the reasonable assumption 
that the source polarization is constant
over the relatively narrow band of wavelengths being considered
for each source.
The quantity of interest is then the change in relative phase 
$\de\ph=4\pi\si\left(L/\la-L/\la_0\right)$
between a wavelength $\la$ and 
a reference wavelength $\la_0$.

The rotation of the Stokes vector
can be expressed via a Mueller matrix $m^{jk}$,
often used to describe the effects 
of filters and polarizers on light
\cite{bw}.
The change in the Stokes vector is given by 
$s^j(\ps,\ch)= m^{jk}(\de\ph) s^k(\ps_0,\ch_0)$,
where $\ps_0$, $\ch_0$ are reference polarization angles.
Some algebra reveals that 
\widetext
\beq
m^{jk}(\de\ph)=\left(
\begin{array}{ccc}
\cxs+\sxs\cdp & \quad \sx\cx(1-\cdp) & \quad \sx\sdp \\
\sx\cx(1-\cdp) & \quad \sxs+\cxs\cdp & \quad -\cx\sdp \\
-\sx\sdp & \quad \cx\sdp & \quad \cdp
\end{array}
\right). 
\label{mueller}
\eeq
\bottom{-2.7cm}
\narrowtext
\noindent

The polarization angle $\ch$,
which measures the amount of circular polarization,
is unavailable in most published literature
on astronomical spectropolarimetry.
We therefore focus attention here on the change 
$\de\psi=\psi-\psi_0$ in the polarization angle $\ps$.
After some algebra, 
we find 
\beq
\de\psi
=\half\tan^{-1}{\fr
{\sin\tilde\xi\cos\ze_0+\cos\tilde\xi\sin\ze_0\cos(\de\ph-\ph_0)}
{\cos\tilde\xi\cos\ze_0-\sin\tilde\xi\sin\ze_0\cos(\de\ph-\ph_0)}},
\label{dpsi}
\eeq
where $\tilde\xi=\xi-2\psi_0$ and 
$\ph_0\equiv \tan^{-1}(\tan2\ch_0/\sin\tilde\xi)$,
$\ze_0\equiv \cos^{-1}(\cos2\ch_0\cos\tilde\xi)$.

To obtain a bound on the coefficients $\kfi$,
our strategy is to fit Eq.\ \rf{dpsi} to
polarization measurements of distant sources at multiple wavelengths.
Since a single source only constrains $\si$ and
$\tilde\xi$ in the particular primed frame appropriate for that one source, 
a measurement involves only a two-dimensional surface
in the ten-dimensional coefficient space.
It follows that at least five different sources are
required to make a complete measurement of all the coefficients $k^a$.

\begin{center}
\begin{tabular}{|l||c|c|c|c|}
\hline
\multicolumn{1}{|c||}{Source} 
& $L$~(Gpc) & $10^{30}L/\la$ & $\log_{10}\si$ & Ref. \\ 
\hline \hline
IC 5063                & 0.04 & 0.56 - 2.8 & -30.8 & \cite{hough} \\
3A 0557-383            & 0.12 & 2.2 - 8.4  & -31.2 & \cite{brindle} \\
IRAS 18325-5925        & 0.07 & 1.0 - 4.9  & -31.0 & \cite{brindle} \\
IRAS 19580-1818        & 0.13 & 1.8 -  9.1 & -31.0 & \cite{brindle} \\
3C 324                 & 1.69 & 58 - 130   & -32.2 & \cite{cimatti465} \\
3C 256                 & 1.92 & 70 -  140  & -32.2 & \cite{dey} \\
3C 356                 & 1.62 & 57 - 120   & -32.2 & \cite{cimatti476} \\
F J084044.5+363328 & 1.71 & 62 - 120   & -32.2 & \cite{brothertonfirst} \\
F J155633.8+351758 & 1.82 & 67 - 110   & -32.2 & \cite{brothertonfirst} \\
3CR 68.1               & 1.70 & 59 - 130   & -32.2 & \cite{brotherton} \\
QSO J2359-1241         & 1.48 & 87 - 90    & -31.1 & \cite{brothertonqso} \\
3C 234                 & 0.55 & 51 - 75    & -31.7 & \cite{kishimoto} \\
4C 40.36               & 2.02 & 73 - 160   & -32.2 & \cite{vernet} \\
4C 48.48               & 2.04 & 75 - 160   & -32.2 & \cite{vernet} \\
IAU 0211-122           & 2.04 & 74 - 160   & -32.2 & \cite{vernet} \\
IAU 0828+193           & 2.08 & 78 - 160   & -32.2 & \cite{vernet} \\
\hline
\end{tabular}
\end{center}
\begin{center}
Table 1. Source Data. 
\end{center}

Since the sensitivity to $k^a$ is inversely related to wavelength,
polarimetry at shorter wavelengths yields better bounds. 
To maximize the sensitivity,
we restrict attention to a chosen sample of 16 sources
with published polarimetric data well suited to our purpose.
Table 1 lists this sample,
which spans wavelengths in the range 400-2200 nm.
For each source,
we choose $\ps_0$ as the mean polarization angle
and use Eq.\ \rf{dpsi} to create a $\ch^2$ distribution for $\de\ps$ 
as a function of $\si$, $\tilde\xi$, $\la_0$, $\ch_0$.
For each $\si$, $\tilde\xi$,
we minimize with respect to $\la_0$ and $\ch_0$,
which produces values consistent with expectations. 

Figure 1 shows the change in the minimized distributions
from their least value for the source 3CR 68.1.
The contour with $\De\ch^2 \equiv \ch^2-\ch^2_{min}=50$
is displayed in the $\tilde\xi$-$\log\si$ plane. 
This corresponds to confidence level $(100-10^{-9})$\%,
which we regard as sufficient for a definite bound.
Only $-90^\circ\le\tilde\xi\le90^\circ$ 
need be considered,
since $\de\ps$ is symmetric under 
$\tilde\xi\rightarrow\tilde\xi\pm180^\circ$.
The general shape of this plot is common to all sources.
Each source eliminates some region of coefficient space
away from $\si=0$, $\tilde\xi=0,\pm90^\circ$,
which are the only configurations
for which the theory predicts no change in the polarization angle.
In principle,
by combining data from multiple sources
it would be possible to eliminate the regions near 
$\tilde\xi=0,\pm90^\circ$
for all the sources.
However,
in the present context it suffices
to make the reasonable assumption 
that the true signal does not lie in 
these regions for all the sources.
Under this assumption,
we have chosen conservative bounds on $\si$ for each source, 
listed in the fourth column of Table 1.
For the source 3CR 68.1, 
the bound is shown as a horizontal line across Fig.\ 1.

\begin{figure}[p]
\centerline{
\psfig{figure=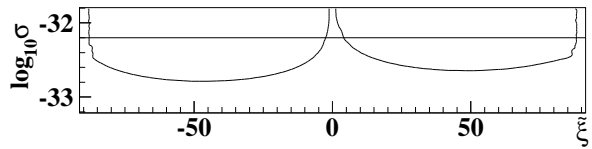,width=0.9\hsize}}
\caption{
Contour with $\De\ch^2=50$ for the source 3CR 68.1.}
\label{fig1}
\end{figure}

To estimate the constraint on $k^a$
that results from combining these bounds,
we assume that the data for each source are consistent with $\si=0$.
The bounds on $\si$ can then be regarded as conservative estimates 
of the error $\de\si$ in a null measurement of $\si$.
We then create a second $\ch^2$ distribution,
$\ch^2=\sum_j{(\si_j)^2}/{(\de\si_j)^2}$,
where $j$ ranges over the 16 sources.
Constant values of this distribution correspond 
to ten-dimensional ellipsoids in the coefficient space.
Taking the magnitude $|k^a|=\sqrt{k^ak^a}$ of $k^a$
as the variable of interest
and minimizing $\ch^2$ with respect
to the other nine degrees of freedom
produces a constraint on $|k^a|$. 
Our conservative estimates yield
$|k^a| < 3\times 10^{-32}$
at the 90\% confidence level.

As a check on this constraint,
we have performed a grid search on the ten-dimensional 
space of coefficients $k^a$.
Fixing the magnitude of the ten-dimensional vector $k^a$ 
in Eq.\ \rf{ka} leaves a nine-dimensional sphere. 
For a given point on the sphere,
$\si$ and $\tilde\xi$ for each source can be found 
and the value of $\ch^2$ obtained using the minimized $\ch^2$ plots.
Summing over sources yields a net $\ch^2$
minimized with respect to all parameters
other than the coefficients $k^a$.  
A grid search on the nine-sphere with points 
separated by about 5 degrees of arc 
was used to find the minimum net $\ch^2$.
Repeating for different values of $|k^a|$
gives results confirming our bound.
To check our procedure directly,
we also generated synthetic polarization data 
with a known faux Lorentz-violating signal
and verified that our procedure correctly extracts it.
Note that the absence of a signal emerging from our analysis
indicates that systematic effects are irrelevant.
To generate a false signal, 
these would need to mimic 
the expected direction and wavelength dependence.

No previous bounds on the coefficients $\kfi$ exist.
However,
constraints on different Lorentz-violating coefficients 
in the fermion sector of the standard-model extension
have been obtained from 
studies of neutral-meson oscillations
\cite{ak,k99,ckpvi,bexpt},
comparative tests in Penning traps \cite{bkr,gg,hd,rm},
spectroscopy of hydrogen and antihydrogen \cite{bkr2,dp},
measurements of muon properties
\cite{bkl,vh},
clock-comparison experiments
\cite{kla,lh,db},
observations of the behavior of a spin-polarized torsion pendulum
\cite{bk,bh},
and studies of the baryon asymmetry \cite{bckp}.
The constraint reported here is comparable to the best 
of these existing limits,
presently a few parts in $10^{31}$.

An improved bound could be obtained 
as more high-quality data become available,
particularly if measurements of the polarization angle $\ch$
could be incorporated in the analysis.
Also, 
the sensitivity to inverse wavelength 
implies that spectropolarimetry of cosmological sources 
at frequencies above the ultraviolet band would be of interest.
The technology to undertake X-ray polarimetry of cosmological sources 
is presently being developed 
\cite{costa},
which suggests an improvement of several orders of magnitude 
may eventually be attainable.

\smallskip
This work was supported in part 
by DOE grant DE-FG02-91ER40661
and by NASA grant NAG8-1770.

\end{multicols}

\begin{references}
\bibitem{fn2}
For a source at declination $d$ and right ascension $r$,
the direction of propagation towards the Earth is  
$\pht^{\,\mu}=(1;-\cd\cor,-\cd\sr,-\sd)$.
The matrix 
$$R^{jk}=\left(
\begin{array}{ccc}
\sd\cor & \sd\sr & -\cd \\
\sr & -\cor & 0 \\
-\cd\cor & -\cd\sr & -\sd
\end{array}
\right)
$$
specifies the rotation between the celestial equatorial and primed frames.
\end{references}
\end{document}